\documentclass[lettersize,journal]{IEEEtran}
\usepackage{amsmath,amsfonts,bm}
\usepackage{algorithmic}
\usepackage{algorithm}
\usepackage{array}
\usepackage[caption=false,font=footnotesize]{subfig}
\usepackage{textcomp}
\usepackage{stfloats}
\usepackage{url}
\usepackage{verbatim}
\usepackage{graphicx}
\usepackage{cite}
\usepackage{tikz,xfp,tkz-fct}
\usepackage{xcolor}
\usepackage{pgfplots}
\usepackage{dsptricks}
\usepackage{circuitikz}
\usepackage[background=none, arrow=red]{callouts}
\usepackage{multirow}
\usetikzlibrary{matrix,babel, arrows.meta, calc,positioning, shapes.misc,graphs,quotes, decorations.pathmorphing, decorations.pathreplacing, decorations.shapes,datavisualization.formats.functions,fit,math,decorations.text}
\DeclareMathSymbol{\shortminus}{\mathbin}{AMSa}{"39}
\begin{document}

%\title{Design and Implementation of Parallel Pipeline FFT Structure for High-speed Real-time Frequency Measurement}
\title{Real-time frequency measurement based on parallel pipeline FFT for time-stretched acquisition system}
\author{Ruiyuan~Ming, Peng~Ye, Kuojun~Yang,~\IEEEmembership{Member,~IEEE}, Zhixiang~Pan, ChenYang~Li, Chuang~Huang%
        % <-this % stops a space
\thanks{P. Ye is with the Faculty of the University of Electronic Science and Technology of China.}% <-this % stops a space
%\thanks{This work was supported in part by the National Natural Science Foundation of China under Grant 61701077 titled “Research
	%	on Key Technology of Broadband Signal Spectrum Analysis”.}
%\thanks{R. Ming, P. Ye and K. Yang are with the School of Automation Engineering, University of Electronic Science and Technology of China, Chengdu 611731, China (e-mail:mingruiyuan@std.uestc.edu.cn).}% <-this % stops a space
%\thanks{This work was supported in part by the National Natural Science Foundation of China under Grant 61701077 titled “Research
%	on Key Technology of Broadband Signal Spectrum Analysis”.}
}


%\markboth{IEEE Transactions on Circuits and Systems I: Regular Papers}{
%	%Ruiyuan Ming, Peng Ye, \MakeLowercase{\textit(et al.)}: 
%	Wideband Sample Rate Converter Using Cascaded Parallel-serial Structure for Synthetic Instrumentation}

\maketitle

\begin{abstract}
Real-time frequency measurement for non-repetitive and statistically rare signals are challenging problems in the electronic measurement area, which places high demands on the bandwidth, sampling rate, data processing and transmission capabilities of the measurement system. The time-stretching sampling system overcomes the bandwidth limitation and sampling rate limitation of electronic digitizers, allowing continuous ultra-high-speed acquisition at refresh rates of billions of frames per second. However, processing the high sampling rate signals of hundreds of GHz is an extremely challenging task, which becomes the bottleneck of the real-time analysis for non-stationary signals. In this work, a real-time frequency measurement system is designed based on a parallel pipelined FFT structure. Tens of FFT channels are pipelined to process the incoming high sampling rate signals in sequence, and a simplified parabola fitting algorithm is implemented in the FFT channel to improve the frequency precision. The frequency results of these FFT channels are reorganized and finally uploaded to an industrial personal computer for visualization and offline data mining. A real-time transmission datapath is designed to provide a high throughput rate transmission, ensuring the frequency results are uploaded without interruption. Several experiments are performed to evaluate the designed real-time frequency measurement system, the input signal has a bandwidth of 4 GHz, and the repetition rate of frames is 22 MHz. Experimental results show that the frequency of the signal can be measured at a high sampling rate of 20 GSPS, and the frequency precision is better than 1 MHz.
\end{abstract}

\begin{IEEEkeywords}
	Frequency measurement, real-time analysis, time-stretched sampling, pipelined fast Fourier transform (FFT), parabolic fitting. 
\end{IEEEkeywords}

\section{Introduction}
\IEEEPARstart{T}{he} real-time measurement of fast non-repetitive events is arguably the most challenging problem in the field of instrumentation and measurement. Observing non-repetitive and statistically rare signals that occur on short timescales requires fast real-time measurements that exceed the speed, precision, and record length of conventional digitizers\cite{Tang1}. A new data acquisition method using Photonic time stretch technology that overcomes the speed limitations of electronic digitizers and enables continuous ultrafast single-shot measurement at refresh rates reaching billions of frames per second with non-stop recording spanning trillions of consecutive frames. Time-stretched sampling has attracted widespread interest in the past decade due to its important potential application, vast of time stretch sampling system has been designed to improve the sampling rate and signal-noise-ratio \cite{yang1,TimeStrechSampling1,TimeStrechSampling2,TimeStrechSampling3,TimeStrechSampling4,TimeStrechSampling5}. Compared with the traditional sampling system, the sampling rate of the time stretch sampling system has increased by tens of times, reaching hundreds of GHz. 

In the high sampling rate time stretched sampling systems, processing high-speed signals in real-time is the bottleneck of signal analysis. In \cite{SpectrumAnalysis1}. A pipelined architecture using a two-level index mapping algorithm for the $N^m$-point $m$-dimension DFT is proposed, this structure is suitable for long sequence analysis, and the resource consumption has been optimized. In \cite{SpectrumAnalysis4}, a pipeline structure R2SD2F architecture based on deep feedback to butterfly-2 is proposed. Compared with other pipeline designs for length-N DFT computation, this structure has fewer complex multipliers and adder consumption, it is efficient for FFT processors in high-precision applications. In \cite{SpectrumAnalysis6}, an interpolated discrete Fourier transformation is proposed, the error of frequency estimation has been reduced, and it has better performance on high-precision frequency estimation. In \cite{SpectrumAnalysis8}, a vibration analyzer based on FFT and periodogram processing is designed to provide an automatic diagnosis of the motor, the designed system computes three FFT, three periodogram estimations, and three postprocessing for diagnosis in about 1.33 ms. This design provides reliable real-time detection. Unfortunately, the response time is not short enough. In \cite{SpectrumAnalysis11}, a reconfigurable real-time spectrum analyzer is implemented on the FPGA platform, an FFT-modified periodogram is used to estimate the spectrum, this design is suitable for cost-sensitive embedded applications but is unable to process high sampling rate signals.  In \cite{SpectrumAnalysis12}, an instantaneous spectral analysis technique based on the extension of Euler’s formula is discussed, this technique is suitable for dealing with nonergodic signal sources having nonstationary spectra, the bandwidth utilized is much smaller than the corresponding one in standard Fourier transform. While the computation complexity is high for long input sequences.  

To analyze the high sampling rate signal at a high repetition rate, a parallel pipelined FFT structure is proposed in this work. A parallel data path structure is used to receive the high sampling rate signals, and a pipelined FFT structure is designed to process these signals at a high repetition rate. Finally, a fitting algorithm is implemented to improve the frequency precision. 

Specially, the contributions of this work are summarized as follows,
\begin{itemize}
	\item The pipeline FFT structure is derived, which supports the real-time spectrum analysis on high speed microwave signals. 
	\item The implementation structure of spectrum fitting is derived, which improve the frequency precision at extremely low cost.
	\item The real-time data transfer structure is presented, which supports high-throughput data stream transferring.
\end{itemize}

The structure of this paper is as follows: the application background is described in Section II. The proposed parallel pipeline FFT structure is elaborated in Section III. Several experimentations are conducted to evaluate performance of the proposed structure in Section IV. Conclusions are finally drawn in Section V.
\section{Background}
\begin{figure}[!t]
	%\begin{figure}[htbp!]
	%	\begin{figure}[!htbp]
		\centering
		\subfloat[]{
			\label{Fig Time Domain}
			\includegraphics[width=3.2in]{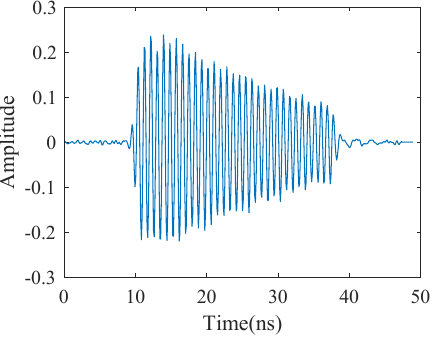}
			}
		%	\quad
		\hfil
		\subfloat[]{
			\label{Fig Frequency Domain}
			\includegraphics[width=3.2in]{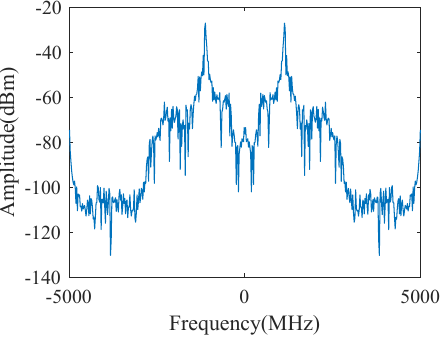}
			}
		\caption[]{Representation of modulated signal: \subref{Fig Time Domain} Time-domain representation of the modulated signal. \subref{Fig Frequency Domain} Frequency-domain representation of the modulated signal.}% (a) Case I. (b) Case II.}
	\label{Fig Pulse}
\end{figure}

The proposed frequency detection system is used to detect a pulsed microwave signal, which is generated by a femtosecond laser-excited frequency-tunable pulsed microwave waveform generator. The applied waveform generator is designed based on an unbalanced single-arm interferometer with frequency-to-time mapping, composed of some easy-to-obtain commercial devices, including mode-locked laser, bandpass optical filter, optical circulator, group delay system, dispersion compensation fiber and photodetector\cite{Tang1}.
The output electrical signal of waveform generator can be expressed as.
\begin{IEEEeqnarray}{rCl}
	I(t) \propto R_e\alpha _R^2 + R_e\alpha _R^2 V cos[2\pi (\Delta t/\beta _2L)t) ]
\end{IEEEeqnarray}
where $R_e$ is the responsivity of the photodetector, $Ar$ is the envelope of the dispersed pulse in the time domain, $V$ is the visibility of the temporal interferogram, $\Delta t$ is the group delay time of the group delay system, $\beta_2$ and $L$ are the group velocity dispersion parameter and length of the dispersion compensation fiber, respectively.

the microwave waveform is shown in Fig. \ref{Fig Pulse}\subref{Fig Time Domain}, the envelope of the microwave waveform is near-triangular-shaped. The pulse length is 45.5 ns, corresponds to the period of pulse. The active signal length is ranging from 26 to 36 ns. The corresponding spectrum is shown in Fig. \ref{Fig Pulse}\subref{Fig Frequency Domain}, it has a relatively narrow center frequency with broadband noise, and the frequency lines have a frequency interval of 22 MHz. 

To detect the main frequency of the pulse microwave in real time, the frequency detection system should meet the following requirements. The supported frequency range should be 100 MHz - 4 GHz, the repetition rate should be higher than 11 MHz, and the frequency accuracy and polarization should be lower than 1 MHz.

\section{Proposed parallel Pipeline Structure}
\subsection{System Structure}
\begin{figure*}[!t]
	%\begin{figure}[htbp!]
	\centering
	\includegraphics{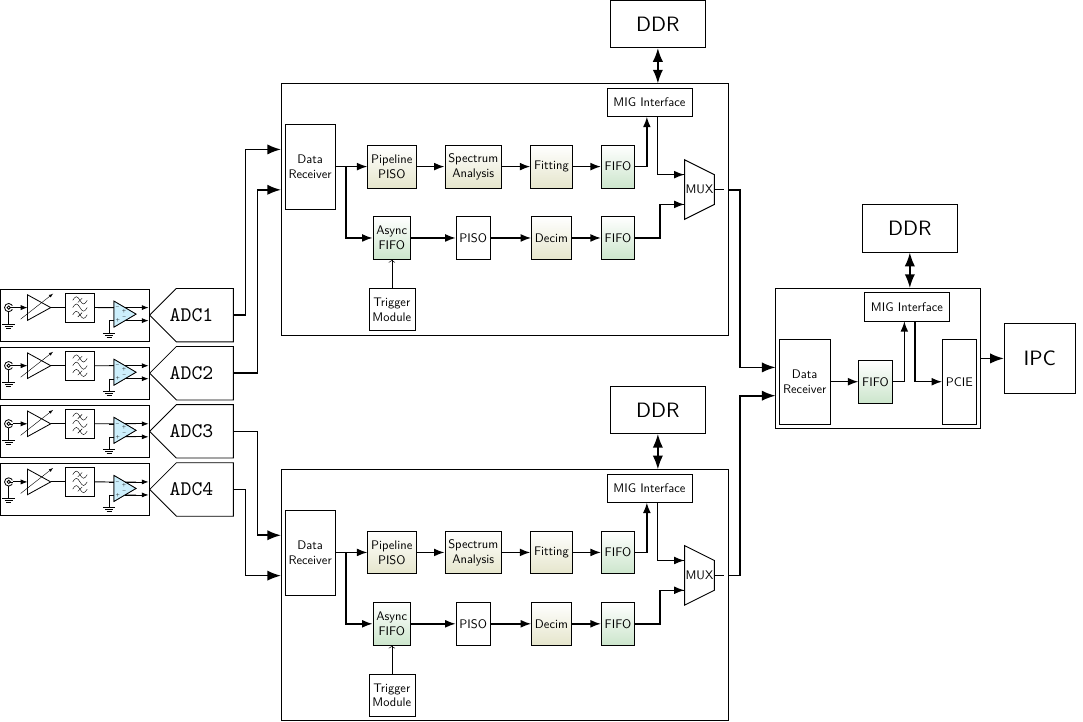}
	\caption{The block diagram of the real-time frequency measurement system, which is composed of signal conditioning, acquisition, processing, and transmission.}
	\label{Fig: System Structure}
\end{figure*}

The real-time frequency measurement system is mainly composed of a signal conditioning module, signal acquisition module, digital signal processing module, data transfer, and industrial personal computer (IPC) module, the block diagram of this system is shown in Fig. \ref{Fig: System Structure}. 

The signal conditioning module adjusts the amplitude of the input signal to an appropriate range. At the same time, it includes the functions of amplification, attenuation, spectrum compensation, and bandwidth limitation. Four condition channels are included to support the simultaneous processing of four sets of signals. 

The signal acquisition module samples the adjusted signal and converts it to a digital signal, the sampling rate of the single channel is 10 GSPS. Four acquisition channels are included, each acquisition channel corresponds to a signal condition channel. 

The digital signal processing module analysis and process signal in the time and frequency domain. It includes two digital paths, the time domain path, and the frequency domain path. The time domain path includes asynchronous FIFO for data reception, parallel-serial conversion for serialization, a decimation module for time resolution adjustment, and synchronous FIFO for data caching. The frequency domain path includes distributed parallel-serial conversion module for real-time serialization, a pipeline FFT module for real-time spectrum analysis, a fitting module for frequency resolution improvement, and a data transfer module for real-time data transfer. The digital signal processing module is implemented in the FPGA of the acquisition card, two signal processing modules are included in this FPGA. Each acquisition card has two acquisition channels. That is to say, each signal processing module corresponds to one acquisition channel. A muxer is used to select the desired path for data transfer. 

The data transfer module is used to transfer the frequency data and waveform data to IPC. When the frequency domain path is selected as the desired path, the data transfer module transfers the frequency result to the IPC in real-time. The bit-width and lane number of the frequency data are firstly adjusted to adapt to the DDR interface, then the adjusted frequency data is stored in the DDR in frames. These frames are uploaded to the IPC through a PCIE interface continuously. When the time domain path is selected as the desired path, the data transfer module transfers the waveform data to the IPC.
In frequency detection mode, the IPC module receives the frequency data and stores these data on a hard disk in real time. In oscilloscope mode, the IPC module receives the waveform data and displays the waveform in real time. In addition, the IPC module receives the user parameter to reconfigure the hardware, including the signal condition parameters, and the signal processing parameters.

The supported frequency range of the frequency detection system is 0 - 4 GHz, the designed signal conditioning channel has a bandwidth of 4 GHz, and the sampling rate of the acquisition system is 10 GSPS to satisfy Nyquist's sampling law. The bit width of ADC is 12 bits to provide high quantization precision and low quantization noise. The capacity of the external DDR for each acquisition board is 4 GByte, and the maximum bandwidth of the DDR interface is 34 GByte/s, which supports the real-time transfer of the frequency data. The transfer rate of the PCIE interface is 5 Gbps, which supports real-time data transfer for four channels.

\subsection{Parallel to Serial module}
\begin{figure}[!t]
	%\begin{figure}[htbp!]
	\centering
	\includegraphics{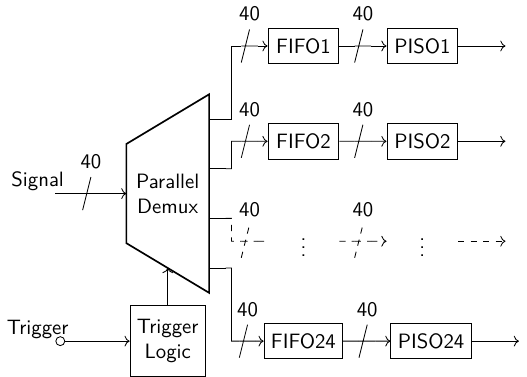}
	\caption{Implementation structure of the typical single-stage P2S module.}
	\label{Fig: OneStage P2S}
\end{figure}

The Parallel-to-Serial (P2S) module is used to convert parallel input data of 40 lanes into 24 groups of serial output, basic units such as demuxer, FIFO, and Parallel Input Serial Output (PISO) modules are usually used to compose the P2S module. The implementation structure of a typical single-level P2S module is shown in Fig. \ref{Fig: OneStage P2S}. The input 40-lane parallel data are distributed to 24-groups FIFO in sequence by demuxer, and converted to serial output by PISO in each group. The trigger logic generated by the external trigger signal controls the time-sharing multiplexing transmission of the parallel demuxer. Each incoming frame is sent to the FIFO in sequence, a new loop will start after the data is sent to the last group of FIFO. 

Single-stage P2S module can realize the conversion operation from 40 lanes parallel inputs to 24 group serial outputs, but the routing of basic units could be tight and the timing closure will be hard to achieve, caused by the high fan-out of parallel demuxer. For the demuxer units, the bit width of the input data is 12bit, parallel lanes are 40, and the group number is 24, so the total output nodes of the parallel demuxer is 11520 bit. Too many nodes lead to higher routing complexity, and the high fan-out and limited routing resources lead to poor timing results. At the same time, the single-stage P2S structure has a low utilization rate of FIFO resources. Since the depth of FIFO can only be set to $2^N$, where $N$ is an integer, and the actual depth required for the data frame is 11, not the power of two. let $N=4$, the utilization rate is $68.75\%$, the total FIFO resource consumed by P2S module is 22.5 kByte. In addition, the conversion from 40 lanes of parallel input to serial output increases the routing complexity of PISO units.

\begin{figure*}[!t]
	%\begin{figure}[htbp!]
	\centering
	\includegraphics{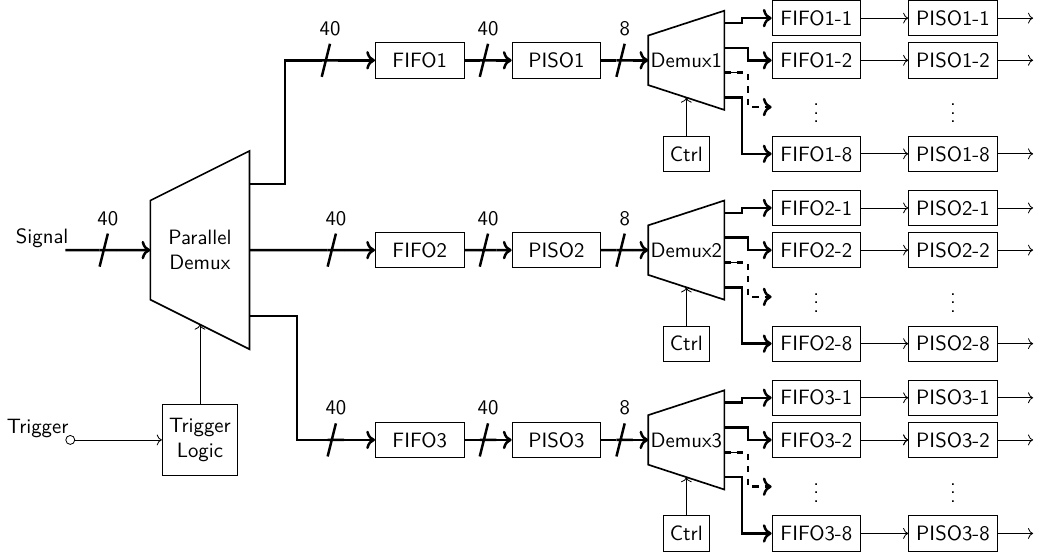}
	\caption{The optimized two-stage P2S structure. The parallel input are converted to serial outputs using cascaded P2S block.}
	\label{Fig: TwoStage P2S}
\end{figure*}

Considering the shortcomings of the single-stage P2S module, an improved two-stage P2S structure is proposed in this paper as shown in Fig. \ref{Fig: TwoStage P2S}. The proposed two-stage P2S module converts the parallel input data to serial output in two steps. Firstly the parallel data are distributed to a few groups, and then each group is distributed to subgroups. The group number is set to 3, subgroup number is set to 8. That is, the first stage distributes a group of parallel data to 3 groups, the input lane number is 40 and the output lane number is 8, the second stage distributes each group to 8 subgroups, the input lane number is 8 and the output lane number is 1. The data of the subgroup is stored in FIFO units and then converted to serial output by the PISO units.
%改进后P2S

In the two-stage P2S structure, the parallel demuxer of the first stage has total output nodes of 1440, $78.5\%$ less than the single-stage P2S structure, which greatly reduces the routing complexity and eventually achieves an optimized timing result. At the same time, the two-stage P2S structure consumes fewer FIFO resources. For the FIFO of the first stage, the utilization rate is the same as the single-stage P2S module, the total consumed FIFO resources is 2.8125 kByte. For the FIFO of the second stage, the input parallel lanes are 8, the required FIFO depth is 55, the final FIFO depth is set to 64, the utilization rate is $85.94\%$, and the consumed FIFO resources is 18 kByte, the total FIFO resources consumed by the two-stage P2S structure is 20.8125 kByte, which consumes $7.5\%$ less FIFO resources than the single-stage P2S structure. In the case that the required FIFO depth is farther from $2^N$, more resources will be saved. In addition, the routing complexity of the PISO units is also reduced. PISO of the first stage achieves a conversion from 40 lanes to 8 lanes, PISO of the second stage achieves a conversion from 8 lanes to a single lane, this reduced conversion ratio results in a low-complexity routing task.

\subsection{Pipeline FFT}
A spectrum analysis module after P2S is required to determine the input signal's frequency. There are 24 groups of serial output from the P2S module, thus 24 lanes spectrum analysis module is required to analyze the respective group. The 24 lanes spectrum analysis modules form a pipeline structure, processing the incoming data frames in sequence.

The frequency detection module is used to detect the main frequency component of the signal and record the adjacent points of the main frequency, and then pass the frequency information to the posterior fitting module in the form of frequency-amplitude pairs. The implementation of frequency detection includes the following steps, padding, FFT, amplitude calculation, and max value selection,  the sub-modules are described in detail below.

\begin{figure*}[!t]
	%\begin{figure}[htbp!]
	\centering
	\includegraphics{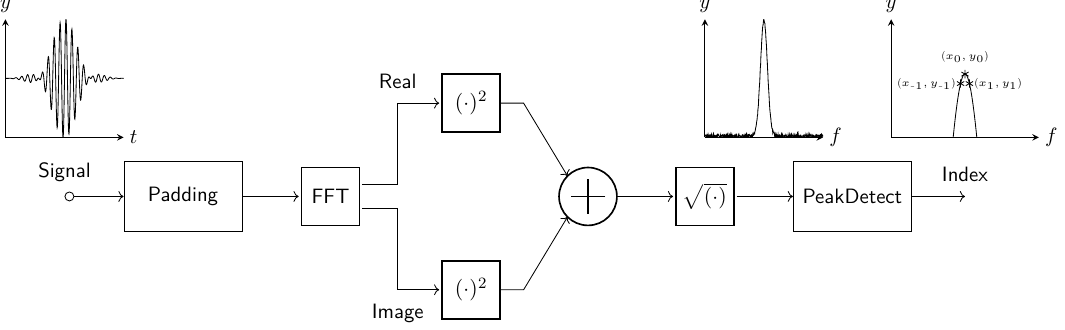}
	\caption{The block diagram of single-lane FFT. The pipelined FFT is composed of tens lanes of FFT units.}
	\label{Fig: FFT}
\end{figure*}

Take the single-lane frequency detection as an example, the block diagram is shown in Fig. \ref{Fig: FFT}. Single input frame is composed of 440 points, the data format is set to S12,0, i.e., signed fixed-point number with a wordlength of 12. The frame is first padded to 512 points to adapt to the posterior FFT core, the padded value is set to 0 to avoid introducing DC bias. The FFT unit conducts the FFT transform on the padded signal to obtain the frequency domain information, the input data bit width and the rotation factor bit width of the FFT are all set to 16 bit to ensure sufficient accuracy of the spectrum, the radix-2 pipeline structure is adopted to support continuous data streams and achieve high efficiency. The output of the FFT unit is complex numbers, which need to be converted into magnitude information. It’s hard to directly calculate the complex magnitude in FPGA, it could be realized by a combination of three basic operations: squaring, summation, and square root operation. Considering that the square root operation only supports floating-point data format, the data sent to the amplitude calculation unit should be converted to floating-point format first. In addition, the floating-point format can achieve higher accuracy and save logic resources. The last step is to detect the main frequency, which is realized by comparing the amplitude at each frequency point in sequences. Then the largest amplitude value and the corresponding frequency index is buffered in registers. At the same time, the amplitude of the two adjacent points of the main frequency point should be recorded, two register pairs are used to store these adjacent points when the main frequency is updated. Considering that the fixed-point number comparison is easier to implement and the latency is smaller than the float-point number comparison, the amplitude data sent to the frequency detection module will be converted to fixed-point format before comparison. The amplitude information recorded by the frequency detection module output to the posterior fitting module in floating-point format to ensure high accuracy. The output results include the amplitude of the main frequency, two adjacent points of the main frequency, index of the main frequency, those are denoted as $y_0$, $y_{-1}$, $y_{1}$, $x_{0}$ respectively.

\subsection{Fitting algorithm}
The fitting module is designed to improve the precision of frequency measurement. Considering that the FFT length of spectrum analysis unit is only 512 points, corresponding spectrum resolution is $F_s/N=20$ MHz, where $F_s$ is the sampling rate of input signal, $N$ is the FFT length. The resolution is too low to meet the measurement requirements. Based on the spectrum characteristics of the modulated signal as described in Section II, the spectrum is composed of two main components, modulation frequency and carrier frequency, and the distance between two frequency points is far, so the influence of energy leakage is trivial. The sideband of the main frequency is mainly determined by the envelope of time series.

\begin{figure}\centering
	\subfloat[]{
		\includegraphics{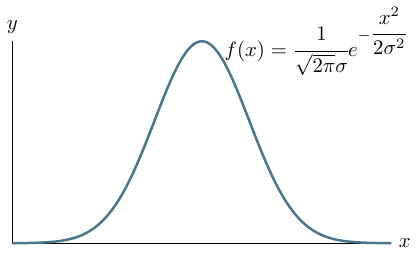}
		\label{fig: Envelope}}
	%	\quad
	\hfil
	\subfloat[]{
		\includegraphics{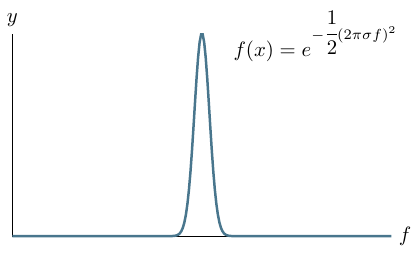}
		\label{fig: Spectrum}}
	\caption[]{Representation of the Guassian curve: \subref{fig: Envelope} Time domain representation of the Guassian curve. \subref{fig: Envelope} Frequency domain representation of the Guassian curve}% (a) Case I. (b) Case II.}
\label{Fig: Gaussian}
\end{figure}

\begin{figure}[!t]
	%\begin{figure}[htbp!]
	\centering
	\includegraphics{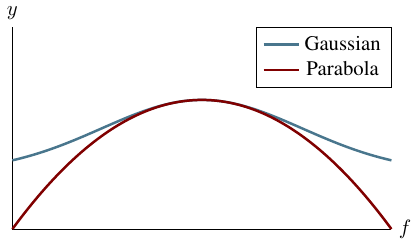}
	\caption{Parabola fitting representation of the Guassian curve in the peak area.}
	\label{Fig: PolyFit}
\end{figure}

Various envelopes have some differences in time domain, but the sideband of the main frequency in frequency domain have similar shapes, and the parabola can be used to fit the sideband to obtain accurate frequency information. Take the Gaussian envelope as an example, the Gaussian curve can be represented as
\begin{IEEEeqnarray}{rCl}
 	f(x)&=&\cfrac {1}{\sqrt{2*\pi}\sigma}e^{-\cfrac{x^2}{2\sigma^2}}
\end{IEEEeqnarray}
where $\sigma$ is the standard variation of Gaussian distribution. The Fourier transform is 
\begin{IEEEeqnarray}{rCl}
	F(w)&=&e^{-\cfrac {(\sigma w)^2}{2}}
\end{IEEEeqnarray}
\cite{book_polyphase1}. the Gaussian curve and its frequency domain representation are shown in Fig. \ref{Fig: Gaussian}. 
A parabolic curve can be used to fit the gaussian curve, the frequency domain representation of gaussian envelope and the fitting parabolic curve are shown in Fig. \ref{Fig: PolyFit}.

Based on the three frequency coordiante captured by the frequency measurement module, the expression of the parabola can be definitely determined, the central frequency point can be derived, the frequency index and amplitude informations of the three coordiante are represented as $(x_{-1},y_{-1})$, $(x_0,y_0)$, $(x_1,y_1)$, the parabolic expression is $y=ax^2+bx+c$. Substitute the coordinate into parabolic expression.
\begin{equation}
	\left\{
	\begin{IEEEeqnarraybox}[\IEEEeqnarraystrutmode
		\IEEEeqnarraystrutsizeadd{1pt}{1pt}][c]{lCl}
		y_{-1} &=& ax_{-1}^2 + bx_{-1} + c	\\	
		y_0 &=& ax_0^2 + bx_0 + c	\\
		y_1 &=& ax_1^2 + bx_1 +c	
	\end{IEEEeqnarraybox}
	\label{Eq Coordinate}
	\right.
\end{equation}
Solving the above equations to obtain the parameters a, b and c, then the center point of parabolic can be represented as $-b/2a$.

\begin{figure}[!t]
	%\begin{figure}[htbp!]
	\centering
	\includegraphics{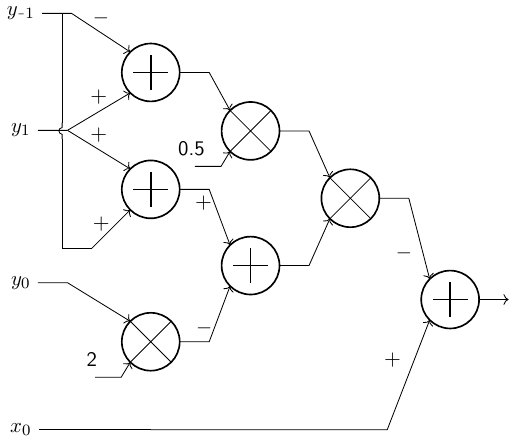}
	\caption{Implementation structure of the pipelined fitting module.}
	\label{Fig: Fit}
\end{figure}

To solve an equation on the FPGA platform is difficult, considering that all frequency indexes are integers, the solving of the equation can be simplified. Represent $x_{-1}$ as $x_0-1$, and $x_1 = x_0+1$, shift the parabolic curve left by $x_0$ units, the new coordinate is 
\begin{equation}
	\left\{
	\begin{IEEEeqnarraybox}[\IEEEeqnarraystrutmode
		\IEEEeqnarraystrutsizeadd{1pt}{1pt}][c]{lClCl}
		x'_{-1} &=& x_{-1}-x_0 &=& -1	\\
		x'_0 &=& x_0-x_0 &=& 0		\\
		x'_1 &=& x_1-x_0 &=& 1	
	\end{IEEEeqnarraybox}
	\label{Eq: tenary}
	\right.
\end{equation}
The shifted curve can be represented as 
\begin{IEEEeqnarray}{rCl}
	y &=& ax'^2 + bx '+ c 		\nonumber\\
	&=& a(x-x_0)^2 + b(x-x_0)+c
\end{IEEEeqnarray}
Substituting the coordinates in (\ref{Eq Coordinate}) into above expression
\begin{equation}
	\left\{
	\begin{IEEEeqnarraybox}[\IEEEeqnarraystrutmode
		\IEEEeqnarraystrutsizeadd{1pt}{1pt}][c]{lCl}
		y_{-1} &=& a -b +c		\\
		y_0 &=& c				\\
		y_1 &=& a + b + c
	\end{IEEEeqnarraybox}
	\label{Eq: tenary}
	\right.
\end{equation}
A and b can be derived
\begin{equation}
	\left\{
	\begin{IEEEeqnarraybox}[\IEEEeqnarraystrutmode
		\IEEEeqnarraystrutsizeadd{1pt}{1pt}][c]{lCl}
		a &=& ((y_1 + y_{-1}) - 2y_0)/2	\\
		b &=& ((y_1 - y_{-1})-2y_0)/2
	\end{IEEEeqnarraybox}
	\label{Eq: tenary}
	\right.
\end{equation}
The center point of the shifted curve is $-b/2a$, thus the center point of the parabola before shifting is
\begin{IEEEeqnarray}{rCl}
x_c &=& -b/2a+x_0	\nonumber\\
	&=& x_0-(y_1-y_{-1}-2y_0)/(y_1+y_{-1}-2y_0)
\end{IEEEeqnarray}
The implementation structure of the fitting module derived from the above expression is shown in Fig. \ref{Fig: Fit}. The fitting input is the coordinates of the three frequency coordinate, and the output is the center frequency point obtained by the fitting module. Firstly, a and b are obtained based on the three frequency coordinates, then the relative position of the center point $-b/2a$ is derived, and finally the center frequency point are obtained by shifted the curve back. Considering that the operation of multiplying by 2 or 0.5 can be implemented by shift operations, this structure actually consumes only four adders and one multiplier.

In the next step, frequency index obtained by fitting process is used to calcutate the actual frequency result, this operation can be implemented by a multiplier. The output data format is set to $S 16,4$, i.e., signed fixed point number with a wordlength of 16 and a fraction length of 4.
\begin{IEEEeqnarray}{rCl}
	f_{mod} = \cfrac {x_c*F_s}{N}
\end{IEEEeqnarray}

\subsection{Real-time Transfer}
The real-time transfer module is designed to transfer the frequency data to IPC in real time. Firstly, consider the bandwidth required for data transmission. As described in Section II, four input channels are realized in this system, the repetition rate of the pulse is 11 MHz, and the frequency data is represented as S16,4, i.e., 16-bit signed fixed point data. So the transmission bandwidth should be 88 Mbyte. The frequency of the modulated signal is first determined based on a digital signal processing algorithm in FPGA to generate frequency data, then this frequency data will be uploaded to an industrial personal computer(IPC) through the PCI-E interface, and finally written to disk by IPC in real-time. Considering that the FPGA device works at a 250 MHz frequency rate, and the bandwidth of the most critical data path is nearly 500 Mbyte/s, the data path in FPGA is sufficient. The PCIE interface is PCIE 2.0 version, the ideal transmission bandwidth is 5 Gbps, and the actual bandwidth is 500 MByte/s, so the PCIE interface also meets the bandwidth requirements of data transmission.
\begin{figure*}[!t]
	%\begin{figure}[htbp!]
	\centering
	\includegraphics{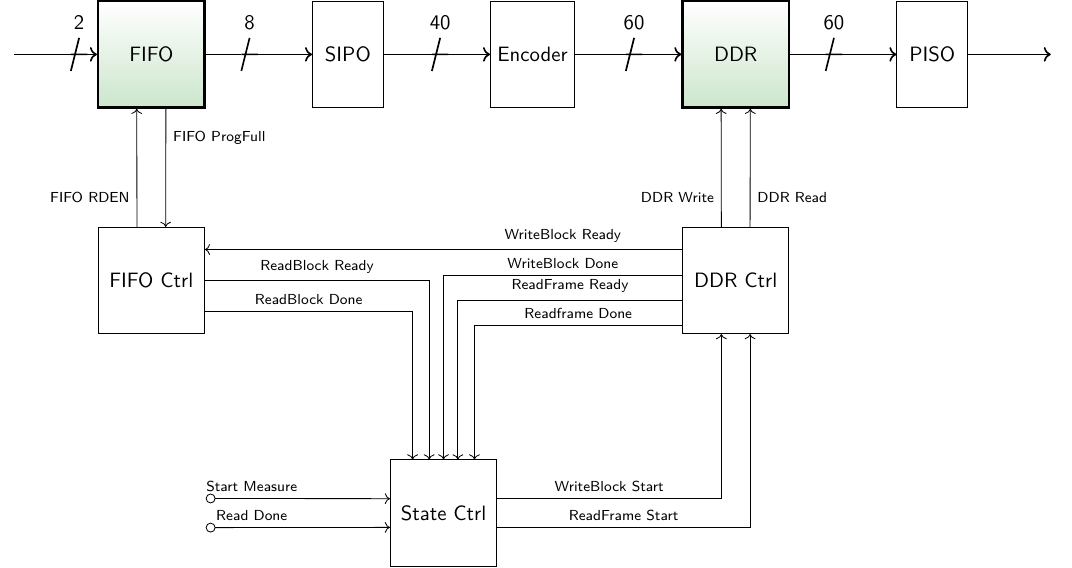}
	\caption{Block diagram of the real-time data transmission module. Two-level caches are used in this structure, the first level is the FIFO unit, the second level is the DDR unit.}
	\label{Fig: RealTime Transfer Structure}
\end{figure*}

There are three critical issues to be considered to determine the data transmission solution. First, in addition to frequency data reading, the IPC needs to configure and initialize the hardware, including the FPGA device and the PCIE interface. To avoid the loss of data during hardware configuration, the frequency data should be temporarily stored in the RAM cache of PCIE. These cached data will be uploaded after the hardware configuration is finished, and the hardware configuration and data reading process will be executed in turn. The size of the PCIE RAM is set to 2.5 MByte, considering that the rate of frequency data is 88 MByte/s, thus a single round of reading and writing process consumes the time of 28.4 ms. According to the actual evaluation, the time for hardware configuration is about 16 ms, and the time of data reading should be less than 12 ms. Secondly, considering the instability of the software operating system, the time consumed in reading data fluctuates. To read a data frame of 2.5 MByte, the average time consumed is about 5 ms, but occasionally there is a transmission timeout. The probability that the reading time exceeds 12 ms is $3\textperthousand$, and there is a 3 in 100000 chance that the time exceeds 100 ms. To avoid the loss of data due to PCIE transmission timeout, a large capacity DDR needs to be arranged before the PCIE stage to store the data during transmission timeout, the capacity of the DDR applied in this design is 2 GByte, which can overcome an extreme timeout of 47 seconds. Finally, the DDR does not support simultaneous read and write operations, when reading the data in DDR, the newly generated frequency data need to be temporarily stored in a FIFO that supports simultaneous read and write operations, and the size of the FIFO should be determined based on the data reading time. Considering that the transmission bandwidth of DDR is 500 MByte/s, the time consumed to transfer 2.5 MByte data is 5 ms, and the data size to be cached is 440 kByte. In other words, the FIFO capacity should be more than 440 kByte. 

The real-time transfer structure consisting of FIFO, DDR, and PCIE is illustrated in Fig. \ref{Fig: RealTime Transfer Structure}. The FIFO is served as the first level cache to store the frequency data when DDR flushes out data, the data size in a single transfer from FIFO is described as a block, one Block is 80 k points, that is 160 kByte. DDR is used as the second level cache to store the frequency data when PCIE RAM flushes out data, the data size in a single transfer from DDR is described as a frame, one frame is composed of 20 blocks, that is 2.5 MByte. The frequency data are first stored in the FIFO, whenever the FIFO threshold of one block is satisfied, the FIFO data will be automatically moved to the DDR. The data from FIFO will go through the SIPO and Encoder modules to adapt to the bit-width of the DDR interface. State Ctrl module is used to receive the status of FIFO and DDR from the FIFO Ctrl module and DDR Ctrl module, simultaneously receive the start measurement command and data read completion flags from the IPC, which is used to determine the data transfer process, such as reading block from FIFO and flushing out a frame from DDR. The FIFO Ctrl module detects the FIFO status and sends these statuses to the State ctrl module, simultaneously receiving the  \textsf{DDR\_write\_block\_ready} command to launch a new block transfer. The DDR ctrl module detects the status of DDR and sends these statuses to the State ctrl module, and receives the  \textsf{DDR\_read\_frame\_start} command to launch a new frame transfer. The DDR write address starts from the initial address and accumulates continuously along with the write block operation, a new cycle will start after the end address is reached. The DDR read address starts from the initial address and accumulates continuously along with the read frame operation, a new cycle will start after reaching the end address. When the read address catches up with the write address, the process waits for the block writing operation before starting a frame reading operation. When the PCIE reading timeout causes the write address to be much larger than the read address, several frames will be flushed out immediately to balance the read and write addresses.
\begin{figure*}[!t]
	%\begin{figure}[htbp!]
	\centering
	\includegraphics{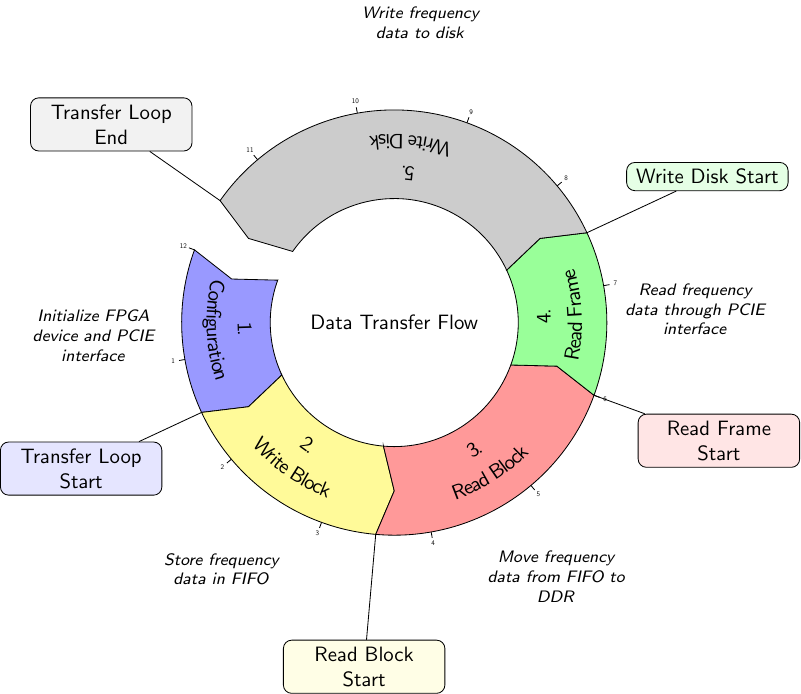}
	\caption{The contrl flow of the real time data transmission. Switch the write/read operation of the DDR unit, store the continuous data in FIFO unit when DDR is in read state.}
	\label{Fig: RealTime Transfer Flow}
\end{figure*}

The data transfer loop is shown in Fig. \ref{Fig: RealTime Transfer Flow}. Before the data transfer begins, the DDR and PCIE are initialized, and the frequency data is stored in the FIFO at the same time. When the FIFO contains a block of data, the read block process is started to move the frequency data from FIFO to the DDR. When the DDR contains a frame of data, i.e., 20 blocks, the read frame process is started. The frequency data in DDR is read to the IPC memory space through the PCI-E interface. After the frame read is finished, the write disk process is started to store the frequency data in the disk of IPC.

\section{Simulation}
In this section, the range and accuracy of the frequency measurement module are evaluated based on MATLAB simulation. Firstly, the modulated signal is split into frames, then the spectrums of these frames are analyzed, and the peak frequency is fitted with a parabolic fitting algorithm. Finally, the range and accuracy of the frequency results are evaluated.

\begin{figure}[!t]
%\begin{figure}[htbp!]
%	\begin{figure}[!htbp]
	\centering
	\includegraphics[width=3.2in]{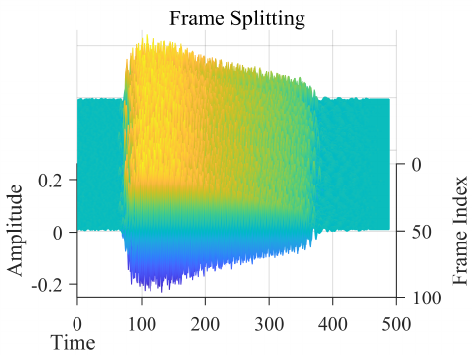}
	\caption{3D plot of the separated frames. The axis x is the time index, axis y is the amplitude, axis z is the frame index.}
	\label{FrameSplinting}
\end{figure}

Firstly, the range and accuracy of a single-tone carrier signal are evaluated. Considering that the length of FFT applied in the implementation structure is 512 points, the acquired signal should be split into consistent frames which have the same length as FFT units. The autocorrelation algorithm is adopted to split the modulated signal to ensure that the cut point is in the gap between two data frames, which avoids potential spectrum leakage. A set of frames are shown in Fig. \ref{FrameSplinting}. The length of these frames is inconsistent and typically less than 512 points, so it is necessary to pad the frame to the length of the FFT unit. Spectrum analysis is performed on these frames and the peak frequency is estimated based on the parabolic fitting algorithm, the frequency accuracy of 512-point FFT reaches 500 kHz, which is approximately the resolution bandwidth of 8k FFT. The range of results is defined as the lowest point of data subtracted from the highest point of data. In this simulation, the range of the fitted frequency results is about 1 MHz.

\begin{figure}[!t]
%\begin{figure}[htbp!]
%	\begin{figure}[!htbp]
	\centering
	\subfloat[]{
		\includegraphics[width=3.2in]{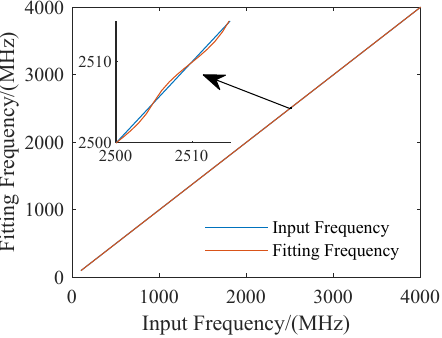}
		\label{Fig: Fitting Frequency}}
	%	\quad
	\hfil
	\subfloat[]{
		\includegraphics[width=3.2in]{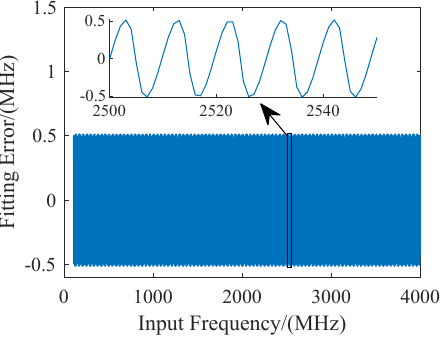}
		\label{Fig: Fitting Error}}
	\caption[]{Precision of the fitting module: \subref{Fig: Fitting Frequency} The fitting result of red line is approximate to the actual frequency of blue line. \subref{Fig: Fitting Error} The fitting error is nearly 0.5 MHz.}% (a) Case I. (b) Case II.}
	\label{Fig Fitting Estimation}
	
\end{figure}

To further evaluate the frequency accuracy and range index of the frequency measurement module in the full bandwidth, a set of modulated signals are generated, which have a different carrier frequency range from 100 MHz to 4 GHz, and the sweep step is 1 MHz. The spectrum analysis and fitting operation are performed on each set of signals. The frequency results of the swept signal sets are shown in Fig. \ref{Fig Fitting Estimation}\subref{Fig: Fitting Frequency}. The deviation of the frequency results from the actual frequency value is shown in Fig. \ref{Fig Fitting Estimation}\subref{Fig: Fitting Error}. The axis x is the sweep frequency, axis y is the frequency result. The maximum deviation of frequency results is 500 kHz, the maximum range is about 1 MHz.

%% scan frequency from 100MHz to 4GHz, 精度和极差评估
\section{Experimentation}
In this section, several experiments are conducted to evaluate the performance of the frequency measurement system. First, the range and precision of the measurement results are estimated based on a modulated signal. Then, a sine sweep test is conducted to evaluate the frequency error over the full bandwidth range. Finally, the influence of different factors such as signal-to-noise ratio (SNR), and amplitude on measurement errors is analyzed.
\subsection{Verification Platform}
\begin{figure}[!t]
	%\begin{figure}[htbp!]
	%	\begin{figure}[!htbp]
		\centering
		\begin{tikzpicture} 
			\node [inner sep=0pt] (brd) at (0mm,0mm) {\includegraphics[width=3.5in]{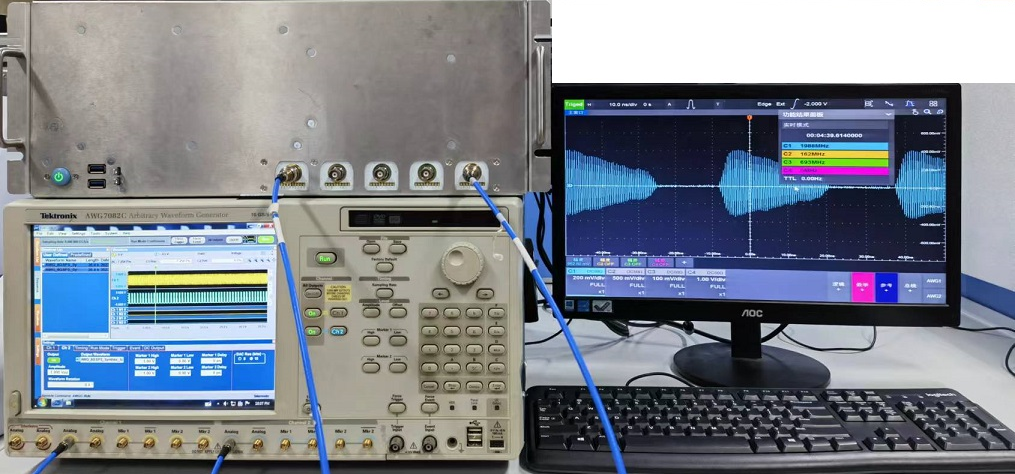}};
			%			\node [inner sep=0pt] (awg) at (-35mm,25mm) {\includegraphics[width=1.5in]{Figure_bitmap/ming6.png}};
			%			\draw [red, thick, ->] (awg) |- node [above right] {} (brd);
			\callout{-16mm,-30mm}{\textbf{Arbitrary Waveform Generator}}{-22mm,-20mm}
			\callout{20mm,25mm}{\textbf{Frequency Measurement Instrument}}{3.5mm,20mm}
			\callout{20mm,-30mm}{\textbf{User Interface}}{21mm,-8mm}
			%			\callout{-30mm,-25mm}{\textbf{Signal Input}}{-25mm,-10mm}
			%			\draw [red, thick] (-21.5mm,-18mm) rectangle (-7mm,10mm);
			%			\draw [red, thick] (-1.5mm,-8mm) rectangle (9mm,3.5mm);
			
		\end{tikzpicture}
		\caption{The verification platform used to verify the performance of the frequency measurement system.}
		\label{Fig: Platform}
	\end{figure}
The performance of the frequency measurement system is evaluated on a hardware verification platform. The applied verification platform is shown in Fig. \ref{Fig: Platform}. The bottom left of the figure is an arbitrary waveform generator (AWG7082C, Tektronix), which is used to generate the modulated signal, supporting a high sampling rate of 20 GSPS. The output frequency of the AWG is ranging from DC to 8 GHz. The upper left is the designed frequency measurement instrument. In this design, the parallel pipelined FFT structure is implemented on a FPGA (Xilinx, XCKU115) device to support high repetition rate frequency measurement. The designed instrument supports a wideband input ranging from DC to 4 GHz. First, the input signals are sampled and quantized at a high sampling rate of 20 GSPS in the embedded acquisition system. Then the sampled signals are processed in the parallel FFT channels to obtained the frequency result. Finally, the frequency results are uploaded to an IPC in real-time for the offline processing and data mining. The right side of the figure is the visualization interface of the IPC. The modulated signal and the frequency results will be visualized on the screen in real-time to achieve accurate and reliable measurements.
\subsubsection{Experiment 1: Frequency}
In this experiment, the precision and range of the frequency measurement system are evaluated. The AWG is used to generate the modulated signal, the carrier frequency is set to 2GHz, and the repetition rate is set to 22 MHz. The frequency measurement system is used to measure the frequency of the modulated signal in real-time, and the generated frequency results are analyzed in MATLAB to evaluate the performance of frequency measurement. Fig. x shows the distribution of the frequency result. As shown in the figure, the frequency of the input signal is 2 GHz, the measured frequency results are ranging from 1.996 GHz to 1.999 GHz.The range of the measurement is 3 MHz, and the precision of the frequency results is nearly 3 MHz, 90\% of the frequency results are concentrated on the 1.998 GHz.
\subsubsection{Experiment 2: Amplitude}
In this experiment, the precision and range of the frequency measurement system are evaluated under multiple amplitude levels. The AWG7082C supports amplitude adjustment in the range of 500 mV - 1 Vpp, and three amplitude configurations (500 mV, 750 mV, and 1 Vpp) are chosen in this experiment. Considering that the input range of ADC is 400 mV in the acquisition system, the modulated signal needs to be attenuated to adapt to the ADC range. The signal conditioning module is used to adjust the attenuation on the modulated signal. An attenuation of 6 dB is chosen for all three amplitude configurations in this experiment, corresponding to the 200 mV amplitude grade of the frequency measurement system. For each amplitude configuration, the frequency range under different carrier frequencies is estimated, the carrier frequencies range from 100 MHz to 4 GHz, and the sweep interval is 1 MHz. The estimation results of the frequency range corresponding to three amplitude configurations are shown in Figure x. In the full bandwidth, the frequency range is within 5 MHz. As the amplitude decreases, the frequency range increases gradually, and the maximum value reaches 4 MHz. The amplitude influences the frequency range due to quantization noise. In the case that the amplitude of the input signal is low, the SNR of the sampled signal decreases considering that the level of quantization noise is constant, which eventually leads to a decrease in the frequency accuracy. With the increase of frequency, the range increases gradually, but it decreases in some frequency points, corresponding to the middle position of the frequency index under 8k FFT, which is consistent with the simulation results in Section IV.

The gain and attenuation of the signal conditioning channel can be adjusted according to the amplitude of the modulated signal during measurement, to achieve higher accuracy and lower range.
\begin{figure}[!t]
	%\begin{figure}[htbp!]
	\centering
	\includegraphics{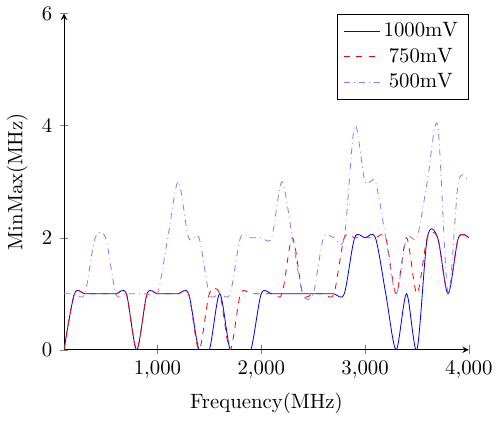}
	\caption{Range Error of the frequency measurement system.}
	\label{Fig: Experiment Amplitude}
\end{figure}
\subsubsection{Experiment 3: SNR}
In this experiment, the precision and range are evaluated under several SNR settings. The modulated signals with different SNRs are obtained by adding white noise to the generated modulated signals. Three SNR configurations of 60 dB, 40 dB, and 20 dB are chosen in this experiment. For each SNR configuration, the frequency range under different carrier frequencies is estimated, the carrier frequencies range from 100 MHz to 4 GHz. the estimation results of the frequency range corresponding to three SNR configurations are shown in Fig. \ref{Fig: Experiment Amplitude}. In the full bandwidth, the frequency range is within 5 MHz. As the SNR decreases, the frequency range increase gradually, the maximum value is 5 MHz. Under a lower SNR setting, the noise floor and spurious components are higher, and the spectral leakage is serious, resulting in deterioration of the frequency range, it is extremely worst when the SNR degraded to 20 dB.
\subsection{Resource Utilization}
\begin{table*}[!t]
	%\begin{table*}[htbp!]
	\centering
	\caption{Resource Consumption of the designed paralel pipelined FFT module}
	\begin{tabular}{lcccccccccc}
		\hline \hline
		& P2S Module 	& Pipeline FFT 	& Fitting Module    	& Real-time Transfer	&Total	&Utilization\\ \hline
		CLB LUTs 			& $3718$ 		& $130594$		& $5000$		& $2349$	& $141661$		& $21.35\%$\\ 
		CLB Flip-Flops  	& $19704$ 		& $265364$		& $9668$		& $3428$	& $298164$		& $22.47\%$\\ 
		Block RAMs			& $75$  		& $168$			& $1$			& $116.5$	& $360.5$		& $16.69\%$\\ 
		DSPs				& $0$			& $2112$		& $20$			& $0$		& $2132$		& $38.62\%$\\ 
		\hline \hline
	\end{tabular}
	\label{Table: Resource}
\end{table*} 
The frequency measurement module is implemented in FPGA (XCKU115), the resource consumption is shown in TABLE \ref{Table: Resource}. The pipeline FFT module consumes the most resources, there are 24 spectrum analysis units in this module, and the resources consumed by each spectrum analysis unit are 5441 CLB LUTs, 11056 CLB Flip-Flops, 7 BRAMs, and 88 multipliers. The pipeline FFT module has the highest computational complexity and processing rate, leading to the highest resource consumption. The real-time transfer module consumes 116.5 BRAM resources, these RAMs are used in the FIFO Block for large data cache. 
\section{Conclusion}
In this work, a high-speed real-time frequency measurement architecture is designed to support real-time spectrum analysis of digital signals in a high sampling rate time-stretch sampling system. The designed real-time frequency measurement architecture consists of three parts, real-time parallel data distribution, pipelined FFT, and optimized spectrum fitting. The real-time parallel data distribution module rearranges input data of hundreds of lanes into real-time serial data, the pipelined FFT structure performs spectrum analysis on each serial lane in real-time. Finally, the spectrum fitting algorithm is implemented on FPGA to improve the frequency accuracy. The real-time frequency measurement architecture is applied to a 20 GSPS digital acquisition system, the spectrum of continuous modulated signals is detected,  the supported repetition rate is up to 11 MHz and the frequency accuracy is better than 1 MHz.
\section*{Acknowledgments}
The authors would like to thank
% Li Chen at University of Electronic Science and Technology of China for many fruitful discussions, thank 
the reviewers for the time spent and their valuable comments.

\bibliographystyle{IEEEtran}
\bibliography{Reference.bib,Reference_Book.bib,Colleagues.bib,Spectrum_Analysis.bib,Time_stretch_Sampling.bib,Time_Stretch_Spectrum_Analysis.bib}
%\begin{thebibliography}{1}
%\bibitem{ref1}
%{\it{Mathematics Into Type}}. American Mathematical Society. [Online]. Available: https://www.ams.org/arc/styleguide/mit-2.pdf
%\end{thebibliography}

\newpage

% your own custom macro containing the $\backslash${\tt{includegraphics}} command to make things
%\bf{If you include a photo:}\vspace{-33pt}

%\begin{IEEEbiography}[{\includegraphics[width=1in,height=1.25in,clip,keepaspectratio]{fig1}}]{Michael Shell}
%Use $\backslash${\tt{begin\{IEEEbiography\}}} and then for the 1st argument use $\backslash${\tt{includegraphics}} to declare and link the author photo.
%Use the author name as the 3rd argument followed by the biography text.
%\end{IEEEbiography}


\vfill

\end{document}